\documentclass[smallextended]{svjour3}   
\usepackage{graphicx}                                   
\usepackage{mathptmx}                                   
\usepackage{amsmath}                                            
\usepackage{amsfonts}                                   
\usepackage{amssymb}                                            
\usepackage{cite}

\usepackage{marvosym}
\usepackage[letterpaper,top=2.5cm,bottom=2.5cm,margin=2.5cm]{geometry}

\usepackage{color}

\usepackage[breaklinks,colorlinks=true,linkcolor=red,citecolor=blue]{hyperref}

\usepackage[labelformat=empty]{subcaption}

\begin{document}

\title{Universal and approximate relations for the gravitational-wave damping  timescale of $f$-modes in neutron stars}
\titlerunning{Universal and Approximate relations for the gravitational-wave damping timescale}

\author{Georgios Lioutas \and Nikolaos Stergioulas}

\institute{Georgios Lioutas \at
              Arnold Sommerfeld Center for Theoretical Physics, Faculty of Physics, Ludwig-Maximilians-Universit{\"a}t, 80333 M{\"u}nchen, 
              Germany \\ 
              \email{Georgios.Lioutas@physik.uni-muenchen.de}           
           \and
           Nikolaos Stergioulas \at
           Department of Physics, Aristotle University of Thessaloniki, Thessaloniki 54124, Greece \\
}

\date{Received: date / Accepted: date}

\maketitle

\begin{abstract}
Existing estimates of the gravitational-wave damping timescale 
of the dominant
quadrupole oscillation mode in the case of rapidly rotating stars are based
on using
a Newtonian estimate for the energy of the mode, in combination with
the lowest-order post-Newtonian quadrupole formula for estimating the gravitational-wave
luminosity.
We investigate a number of other choices for estimating the gravitational-wave damping timescale in the nonrotating limit and construct a highly accurate, empirically corrected formula that has a maximum relative error of only 3\% with respect to the perturbative result in full general relativity. The expressions involved
are sufficiently general
to be  extended to the case of rapidly
rotating stars. We also present a new higher-order empirical relation for the gravitational-wave damping timescale of quadrupole oscillations that is accurate in the whole range of expected values for the compactness of neutron stars, without the need for involving the moment of inertia.    
\keywords{Gravitational waves \and f-mode \and Damping time \and Approximate relation \and Neutron stars}
\end{abstract}

\section{Introduction}

Fundamental properties of neutron stars (such as their radius and the composition of matter at their center) remain elusive, mainly because of large uncertainties in the determination of their radius through observations in the electromagnetic spectrum (see e.g. 
\cite{OzFre2016,NatStKajSulPut2016,WatAndCol2016}). An alternative approach for measuring neutron star radii and constraining the equation state (EOS) of matter at high densities  is the application of gravitational-wave asteroseismology to either isolated stars 
\cite{AnKok1996,AnKok1998,KokSchm1999}
or binary neutron star merger events \cite{StBaZaJa2011,BaJa2012,BaJaHeSch2012,BaStJa2014,BaSt2015}. The successful detection of gravitational waves from neutron star oscillations will rely on accurate models of the expected waveforms 
(see e.g. \cite{ClBaSt2016,BoChReSaTa2017,YaPaYaLePrYu2017}). A critical input for analytic models is the damping timescale of oscillations due to gravitational-wave emission. While this timescale can be accurately computed for nonrotating stars (e.g. \cite{AnKokSch1995}), in the case of rapidly rotating stars (including differentially rotating merger remnants) only very rough estimates exist (e.g. 
\cite{Shib2005,ShTaUr2005,DoGaKoKr2013,ClBaCaJaPaSt2014}). This is because the quasi-normal modes of rapidly rotating neutron stars have only been computed in either the Cowling appoximation (e.g. 
\cite{DoGaKoKr2013,KaWiKo2010,DoKo2015}) or in the IWM-CFC approximation \cite{Yosh2012} where, in both cases, the imaginary part of the quasinormal mode frequency (describing the gravitational-wave damping timescale) is being neglected, see \cite{FrSt2013b,PaSt2016} for extensive reviews.
Quasi-normal $f-$modes of rapidly rotating stars have also been studied in full general relativity, but in simulations where the total simulation time was long enough to only allow the real part of the frequency to be determined, but too short for the gravitational-wave damping timescale to be extracted (e.g. \cite{FoGoIyMiReSeStSuTo2002,ZiKoScSt2010} ). 

Existing estimates of the gravitational-wave damping timescale of the dominant oscillation modes in the case of rapidly rotating stars are based on using a Newtonian estimate for the energy of the mode, $E$, in combination with the lowest-order post-Newtonian quadrupole formula for estimating the gravitational-wave luminosity, $|dE/dt|$,   (e.g. \cite{DoGaKoKr2013,DoKo2015}).
The gravitational-wave damping timescale is then obtained from $1/\tau_{\rm GW}=(1/2E)| dE/dt|$.

Here, we investigate a number of other choices for estimating $\tau_{\rm GW}$ (focusing on the $l=2$ $f-$mode) and compare their relative accuracy with respect to the exact result in the nonrotating limit. We show that a specific choice of $E$ and $dE/dt$, which was used in \cite{Burg2011}, stands out for its higher accuracy. 
Furthermore, we find an empirical relation that corrects for the difference between the approximate and exact results. This allows us to construct a final, corrected, empirical formula for  $\tau_{\rm
GW}$  that is of very high accuracy. The expression is sufficiently general to be  applied also for rapidly
rotating stars. 

The damping timescale $\tau_{\rm GW}$ for $f-$modes was shown to satisfy an empirical relation (as function of the compactness $M/R)$ when scaled by $M^3/R^4$ in \cite{AnKok1998} (where $M$ is the mass and $R$ is the radius of the star). A higher-order relation, but with a very different scaling (to match the scaling of the real part of the eigenfrequency), was proposed in \cite{TsLe2005}. In addition, a highly accurate relation as function of an effective compatness, $\eta$, that involves the moment of inertia $I$ was shown to exist in \cite{LaLeLi2010}, see also \cite{DoKo2015,ChSoKa2015}. Here, we generalize the expression proposed in \cite{AnKok1998} to higher order, showing that it describes  $\tau_{\rm GW}$ with high accuracy for all values of compactness, in contrast to the relation of \cite{TsLe2005}, which is accurate only in a limited range. This eliminates the need for using the moment of inertia. The gravitational-wave damping timescale $\tau_{\rm GW}$ can be thus described by a highly accurate universal relation that only involves $M$ and $R$.

In the following sections, we briefly describe the theoretical and numerical setup, present numerical
results and compare their accuracy. Next, we construct empirical relations and we conclude with a discussion. We set $c=G=1$, unless otherwise specified.
Greek indices are taken to run from 0 to 3, Latin indices from 1 to 3, and we adopt the standard convention for the summation over repeated indices.
\section{Neutron star oscillations}
\subsection{Linear perturbations of nonrotating stars}
\label{sec:lin}
 Extensive reviews of the formalism for linear oscillations of nonrotating relativitistic stars can be found in \cite{KokSchm1999,FrSt2013b}. We consider a nonrotating star in equilibrium, working in Schwarzschild coordinates $t,r,\theta,\phi$. The matter is taken to be a perfect fluid with stress-energy tensor $T_{\alpha\beta}=(\epsilon+p)u_\alpha u_\beta +pg_{\alpha\beta}$, where $\epsilon$ is the energy density, $p$ is the pressure and $u^\alpha$ is the 4-velocity of the fluid. Due to the symmetries of the background, linear perturbations of an equilibrium model can be 
 decomposed into a sum of independent quasi-normal modes, each depending on a specific spherical harmonic $Y_{lm}$ (for scalar quantities) and  polar or axial vector or tensor spherical harmonics for vectors or tensors, correspondingly. The time dependence of all perturbed quantities is assumed to be harmonic, i.e. proportional to $e^{i\omega t}$, where  
 $\omega$ is the complex eigenfrequency of the quasinormal mode. The latter can be decomposed as \begin{equation}
  \omega = \sigma + \frac{i}{\tau_{\rm GW}},
 \end{equation}
 where $\sigma:=2\pi f$ is the real part of the pulsation frequency and $\tau_{GW}$\ is the gravitational-wave damping timescale.  
 
Choosing to work in the Regge-Wheeler gauge \cite{RW1957}, the full spacetime metric (equilibrium plus polar perturbation) can be written in the form \cite{LinDet1983}\begin{equation}
\begin{aligned}
ds^2 =& - e^{2\nu} \left( 1 + r^l H_0^{lm} Y_{lm} e^{i\omega t} \right) dt^2 - 2i\omega r^{l+1} H_1^{lm} Y_{lm} e^{i\omega t} dt dr \\&
       + e^{2\lambda} \left( 1 - r^l H_0^{lm} Y_{lm} e^{i\omega t} \right) dr^2 
       + r^2 \left(1-r^l K^{lm} Y_{lm} e^{i\omega t} \right) 
       \left( d\theta^2 + \sin^2\theta d\phi^2 \right).
\end{aligned}
\end{equation}
 Here, $\nu(r)$ and $\lambda(r)$ are the unperturbed metric functions, while $H_0(r)$, $H_1(r)$ and $K(r)$ are functions describing polar perturbations of the spacetime in this gauge.

The covariant components of the Lagrangian displacement $\xi^\mu$ of fluid elements can be decomposed as \begin{subequations}
 \begin{align}
  \xi_r  &= e^\lambda r^{l-1} W^{lm} Y_{lm} e^{i\omega t} ,\\
  \xi_\theta  &= - r^l V^{lm} \partial_\theta Y_{lm} e^{i\omega t} ,\\
  \xi_\phi   &= - r^l V^{lm} \partial_\phi Y_{lm} e^{i\omega t},
 \end{align}
\end{subequations}
where $W^{lm}$ and $V^{lm}$ are functions of $r$ only.
The  corresponding Eulerian change in the 4-velocity has spatial components $ \delta u^i =u^t \partial_t \xi^i= i\omega e^{-\nu}\xi^i$ and time-component $\delta u^t = 1/2 (u^t)^3\delta g_{tt} =-e^{-\nu}\delta\nu$.
The Eulerian perturbation in the energy density is decomposed as
\begin{equation}
 \delta \epsilon  = r^l \delta \epsilon^{lm} Y_{lm} e^{i\omega t},
\end{equation}
where $\delta \epsilon^{lm}$ is a function of $r$. 

In 
 \cite{LinDet1985} a new variable is defined as 

 \begin{equation}
  X^{lm} = \omega^2(\epsilon + p)e^{-\nu} V^{lm} - \frac{1}{r}\frac{dp}{dr}
e^{\nu-\lambda} W^{lm} + \frac{e^{\nu}}{2} (\epsilon + p) H^{lm}_0.
 \end{equation}
 and $H_0$ is then eliminated using one of the perturbed field equations.
This yields a system of differential equations for
 the four functions  $H^{lm}_1,K^{lm},W^{lm}$ and $X^{lm}$. 

Requiring regularity
at the center, one obtains boundary conditions for $H_1^{lm}$ and $X^{lm}$. Because $X^{lm}$ is directly related to the Lagrangian perturbation of the pressure,  setting it to zero 
 provides another boundary condition at the surface. The above three conditions suffice to obtain solutions of the perturbed quantities in the interior of the star, for arbitrary values
of the complex frequency $\omega$, corresponding to 
a combination of ingoing and outgoing waves.

 The eigenfrequencies of the (astrophysically relevant) purely outgoing quasi-normal modes form a discrete spectrum of frequencies for which there are no incoming waves. These are obtained by matching the interior solution to the exterior solution described by  the Zerilli
function  \cite{Zer1970}
 \begin{equation}
 Z^{lm} = \frac{r^{l+2}}{nr+3M}(K^{lm}-e^{2\nu}H^{lm}_1),
\end{equation}
 where $n=(l-1)(l+2)/2$, demanding continuity
at the surface as well as continuity of first derivatives. In the numerical implementation, we adopt the approach
of Andersson, Kokkotas and Schutz \cite{AnKokSch1995}, introducing a new
variable $\Psi^{lm}$ in  place of $Z^{lm}$ as
 \begin{equation}
  \Psi^{lm} = \left( 1 - \frac{2M}{r} \right) Z^{lm},
 \end{equation} which has the benefit of suppressing oscillatory behaviour
in the solution. Due to the spherical symmetry of the background, we only consider the $m=0$ case. \subsection{Pulsation energy}
 The energy
in the oscillation mode will be the sum of the kinetic energy and the potential energy, the two energies being out of phase by $\pi/2$ and both decaying with time as $e^{-2t/\tau_{\rm GW}}$. For example, if the kinetic energy is $E_{\rm kin}=E^0_{\rm kin}\sin^2(\sigma t) e^{-2t/\tau_{\rm GW}}$, then the potential energy will be  $E_{\rm
pot}=E^0_{\rm pot} \cos^2(\sigma t)e^{-2t/\tau_{\rm GW}}$, where $E^0_{\rm kin}$ and $E^0_{\rm pot}$ are the maximum values of these energies in the first oscillation cycle. 

The total energy $E_{\rm mode}$ will be $E:=E_{\rm kin}+E_{\rm pot}=E^0_{\rm mode}e^{-2t/\tau_{\rm GW}}$ and since $E^0_{\rm mode}$ should be constant within one oscillation cycle,  it follows that $E^0_{\rm mode}=E^0_{\rm kin}=E^0_{\rm pot}$ \cite{ThoCam1967}. For calculating $E^0_{\rm mode}$ it is thus sufficient to calculate the kinetic energy $E^0_{\rm kin}$ as the integral over the proper volume of the redshifted total kinetic energy
density:
 \begin{eqnarray}
 E^0_{\rm mode}=E^0_{\rm kin} &=&  \int_V \frac{1}{2}(\epsilon+p)\delta v^i \delta v^\ast_i(u^t)^{-1}\sqrt{{}^3g}d^{3}x, \label{enerns1}\\
   &=&\frac{1}{2} \sigma^2 \int_{0}^R dr\; r^{2l} (\epsilon+p) e^{\lambda-\nu}
\left[ |W^{l0}|^2 + l (l+1) |V^{l0}|^2 \right], \label{enerns}
\end{eqnarray}
where  ${}^3g$ is the determinant of the induced metric on a $t-$const. hypersurface, $u^t$ is the $t$-component of the 4-velocity $u^\alpha$ and $\delta v^i
\delta v_i^\ast=g^{ij}\xi_{i,t}\xi^\ast_{j,t}e^{-2\nu}$. The asterisk sign
${}^\ast$ implies complex conjugation.

Notice that instead of the maximum kinetic energy $E^0_{\rm kin}$, one could use the averaged (over a full period) kinetic energy $ <E_{\rm kin}>$, in which case the pulsational energy of the mode would be  $ E_{\rm kin}^0=2< E_{\rm kin}>$, as in
\cite{GuKaChGu2013}. 

\subsection{Gravitational-wave luminosity}

The luminosity in gravitational waves can be estimated using the \textit{standard quadrupole formula} (SQF), which assumes slowly-varying, weak-field sources (see \cite{LanLif1962}). In terms of the reduced quadrupole moment tensor
\begin{equation}
Q^{ij}=\int \rho\left(x^i x^j-\frac{1}{3}r^2\delta^{ij}\right)dV
\end{equation}
 and averaging over several characteristic periods, the quadrupole formula for the rate of energy loss due to gravitional wave emission becomes
\begin{equation}
\left <\frac{dE}{dt} \right >_{\rm GW}=-\frac{1}{5}< \dddot Q_{ij} \dddot
Q_{ij}>
\label{quadformula}
\end{equation}
(e.g. \cite{Maggior2007}). For an $l=2$, $m=0$ mode, the Eulerian perturbation in the rest-mass density
can be decomposed as 
\begin{eqnarray}
\delta \rho &=& \delta \rho(r)Y_{20}  e^{i\omega t} \\
&=&\delta \rho(r) \sqrt{\frac{5}{4\pi}} P_2(\cos{\theta}) e^{i\omega t},
\end{eqnarray}
with the usual definitions of spherical harmonics and Legendre polynomials.
The explicit evaluation of (\ref{quadformula}) then yields (\cite{OsaHan1973})
\begin{equation}\label{quadenns}
 \left <\frac{dE}{dt} \right >_{\rm GW} = -\frac{4 \pi}{75}  \sigma^6 \left( \int_0^R r^4 \delta\rho(r)
dr \right)^2,
 \end{equation}
where $\delta\rho(r)$ is the radial part of the Eulerian perturbation in the Newtonian rest-mass density.

Equation (\ref{quadenns}) is only a lowest-order approximation, whereas the pulsation energy (\ref{enerns}) is known in full general relativity. Combining (\ref{quadenns}) with (\ref{enerns}) to obtain $\tau_{\rm GW}$ is thus inconsistent. In the absence of better choices\footnote{We are interested in obtaining  $\left <\frac{dE}{dt} \right >_{\rm GW}$ through integration over the source, rather than over a sphere at large distance.}, one could thus test    various ad hoc modifications of (\ref{quadenns})
and evaluate their effect on computing $\tau_{\rm GW}$. A similar approach has been followed in proposing various modified forms of the SQF for computing the \textit{amplitude} of gravitational waves (see 
\cite{1990-Blanchet-etal,ShiSek2003,NaFoZaPi2005,2005-Dimmelmeier-etal,Baiotti2009}) and here we extend this approach to modifications of the gravitational-wave \textit{luminosity }(\ref{quadenns}).
Specifically, we replace the rest-mass density $\rho$ by an effective density  $\rho_{\rm eff} $ and evaluate the corresponding Eulerian perturbation  $\delta \rho_{\rm
eff}$. 

Table \ref{tab1} lists various choices we make for $\rho_{\rm eff}$ along with the corresponding $\delta \rho_{\rm
eff}$. In (\ref{quadenns}),  $\delta \rho(r)$ is then replaced by $\delta \rho_{\rm eff}(r)$.
The cases SQF1, SQF2 and SQF3 correspond to the same choices of   $\rho_{\rm eff}$ (keeping the same naming convention) as in \cite{Baiotti2009}. The case where  $\rho_{\rm eff}$ is chosen to be the relativistic energy density $\epsilon$ was proposed in \cite{Burg2011} and here we call it the \textit{relativistic quadrupole formula }(RQF) (it is, of course, only an ad hoc modification of SQF and not a fully relativistic formula).

\begin{table}[h!]
\begin{center}
 {
\renewcommand{\arraystretch}{1.5}
\caption{Different choices for the effective density $\rho_{\rm eff}$ and its Eulerian perturbation $\delta \rho_{\rm eff}$ that are used in (\ref{quadenns}) in place of  $\delta \rho$.}
\label{tab1}      
\begin{tabular}{ccc}
\hline\noalign{\smallskip}
 & $\rho_{\rm eff}$  & $\delta\rho_{\rm eff}$  \\
\noalign{\smallskip}\hline\noalign{\smallskip}
SQF & $\rho$ & $\delta\rho$\\ 
 SQF1& $a^2\sqrt{\gamma}T^{tt}$&$2a\delta a\sqrt{\gamma}T^{tt}+a^2\dfrac{\delta\gamma}{2\sqrt{\gamma}}T^{tt}+a^2\sqrt{\gamma}\delta T^{tt}$\\ 
 SQF2 & $\sqrt{\gamma}W\rho$&$\dfrac{\delta\gamma}{2\sqrt{\gamma}}W\rho + \sqrt{\gamma}\delta W \rho + \sqrt{\gamma} W \delta\rho$\\
 SQF3 & $u^t\rho$&$u^t\delta\rho+\rho\delta u^t$\\
 RQF & $\epsilon$&$\delta\epsilon$\\
\noalign{\smallskip}\hline
\end{tabular}
}
\end{center}
\end{table}
 
 The definitions of $\rho_{\rm eff}$ in Table \ref{tab1}, include quantities from a 3+1 split of spacetime (see e.g. \cite{Alcub2012}), namely the lapse function $a = e^{\nu}$, the Lorentz factor $W = u^ta$ and the determinant $\gamma = \det{|\gamma_{ij}|}$ of the 3-metric $\gamma_{ij}$.

\begin{table}[h!]
\begin{center}
 {
\renewcommand{\arraystretch}{1.5}
\caption{Different ways for constructing approximate formulas for $\tau_{\rm
GW}$. In the first column, N stands for the Newtonian pulsation energy (\ref{pulsnewt})
and R stands for the relativistic pulsation energy (\ref{enerns}). }
\label{tabb2}      
\begin{tabular}{ccc}
\hline\noalign{\smallskip}
  $E^0_{\rm mode}$  & $<dE/dt>_{\rm GW}$ &$\tau_{\rm
GW}$ \\
\noalign{\smallskip}\hline\noalign{\smallskip}
N & SQF & N/SQF \\ 
R & SQF1& R/SQF1 \\
R & SQF2& R/SQF2 \\
R & SQF3& R/SQF3 \\
R & RQF& R/RQF \\
\noalign{\smallskip}\hline
\end{tabular}
}
\end{center}
\end{table}

\subsection{Gravitational-wave damping timescale}

The damping timescale of oscillations due to the emission of gravitational
waves may be obtained as
 \begin{equation}
  \frac{1}{\tau_{\rm GW}} = - \frac{<dE/dt>_{\rm GW}} {2E^0_{\rm mode}},
  \label{1tau}
 \end{equation}
 where $<dE/dt>_{\rm GW}$ is the luminosity in gravitational waves, averaged
over a few oscillation periods.

When accurate values for $\tau_{\rm GW}$ are not available (as is currently the case
for rapidly rotating stars), relation (\ref{1tau}) can be used to obtain
an estimate, by choosing suitable expressions for $E^0_{\rm mode}$ and $<dE/dt>_{\rm
GW}$. Table \ref{tabb2} shows the different combinations of  $E^0_{\rm mode}$ and $<dE/dt>_{\rm GW}$ that we consider here.
The SQF is combined only with the Newtonian expression of the pulsation energy (\cite{Thorne1969})
\begin{equation}
  (E^0_{\rm mode})_{\rm Newt}=\frac{1}{2} \int_V  \rho \delta v^i \delta^\ast
v_i   dV, \label{pulsnewt}
 \end{equation} 
 whereas SQF1, SQF2, SQF3 and RQF are combined with the relativistic pulsation energy (\ref{enerns}). The corresponding naming convenction for each combination is shown in the third column of Table \ref{tabb2}.

\begin{table}[h!]
\begin{center}
 {
\renewcommand{\arraystretch}{1.5}
\caption{List of EOS used for constructing nonrotating equilibrium models. }
\label{tabb3}      
\begin{tabular}{cc}
\hline\noalign{\smallskip}

EoS & Reference   \\

\noalign{\smallskip}\hline\noalign{\smallskip}
 WFF1 & Wiringa et al. \cite{eosWFF} (denoted AV14+UVII) \\
 WFF2 & Wiringa et al. \cite{eosWFF} (denoted UV14+UVII) \\
 HHJ & Heiselberg \& Hjorth-Jensen  \cite{eosHHJ1,eosHHJ2}\\
  SkI4 & Chabanat et al. \cite{eosSkyrme1} ; Farine et al. \cite{eosSkyrme2}\\
 Ska & Chabanat et al. \cite{eosSkyrme1} ; Farine et al. \cite{eosSkyrme2}\\
 MDI & Prakash et al. \cite{eosMDI}\\
  O & Bowers et al. \cite{eosO}\\
   N & Walecka \cite{eosN}\\
 L & Pandharipande \& Smith \cite{eosL}\\
\noalign{\smallskip}\hline

\end{tabular}
}
\end{center}
\end{table}

On the other hand, the numerical data for  $\tau_{\rm
GW }$ in full general relativity, as obtained by solving for the complex eigenvalues of the system of perturbation equations described in Section \ref{sec:lin}, will be denoted as perturbative GR in the following Sections.

 \section{Equilibrium models}
\begin{figure}[h]
\centering
  \includegraphics[scale=0.6]{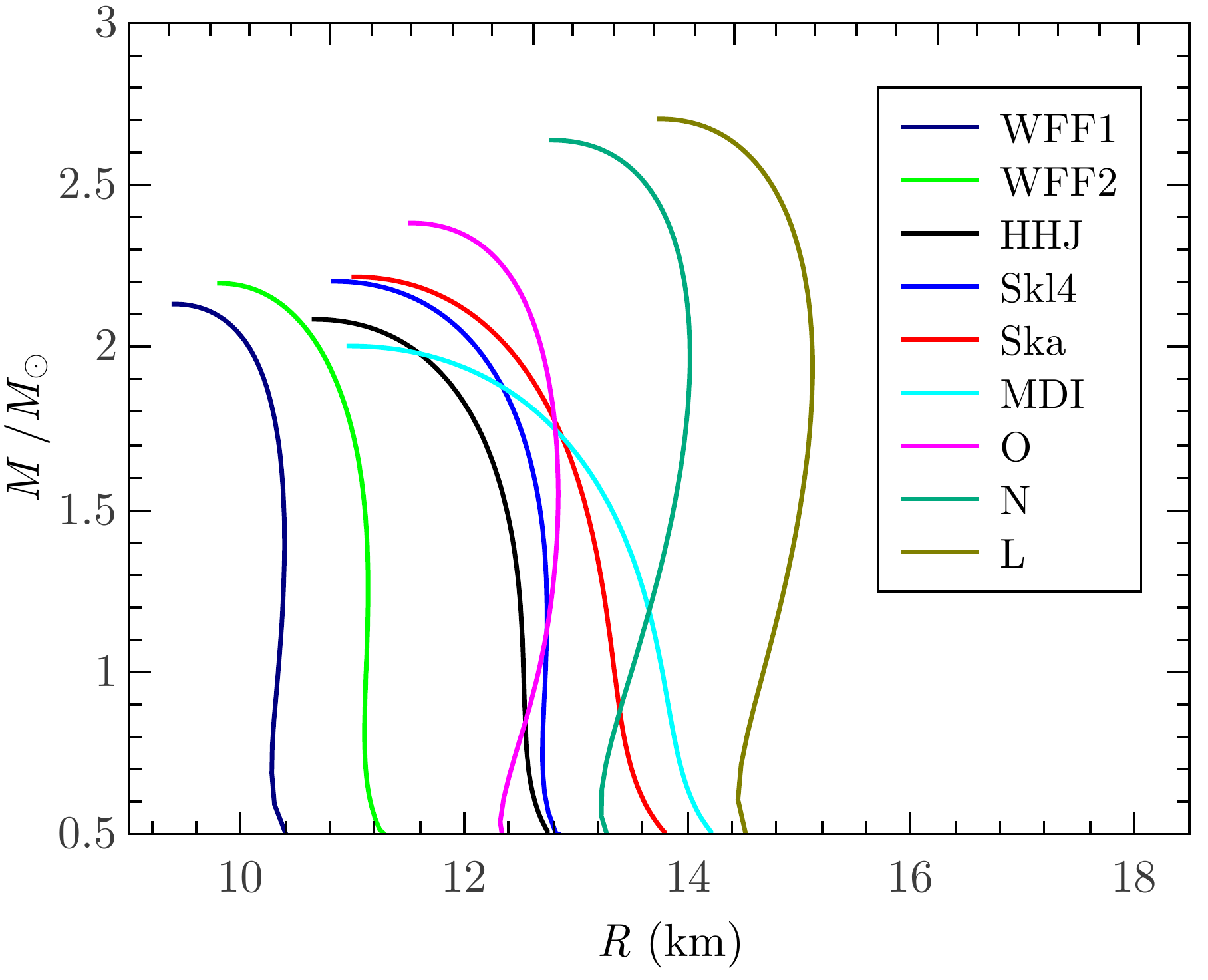}
\caption{Mass-radius relation for the EOS under consideration.}
\label{fig:1}       
\end{figure}

 We use nine different EOS, which cover a large part of the allowed mass-radius parameter space, while satisfying 
 the observational constraint of producing a maximum-mass configuration with gravitational mass $M>2M_\odot$.  Table \ref{tabb3} lists the main reference for each EOS (WFF1, WFF2, SkI4, Ska, MDI(L95), H-HJ(d=0.13), O, N, L).
Figure \ref{fig:1} displays the gravitational mass vs. radius relation for models of nonrotating neutron stars constructed with the above nine EOS.
Only stable models, up to the maximum mass, are considered.  The maximum mass ranges between 2$M_\odot$ and 2.7$M_\odot$,whereas typical radii are in the range between $\sim 10$km for very soft EOS and $\sim 15$km for very stiff EOS, which is a measure of the current uncertainty in the theoretical description of the EOS matter at densities larger than a few times nuclear density. 

  \section{Universal relation for the gravitational-wave damping timescale}
  

  \begin{figure}[h]
  \centering
  \includegraphics[scale=0.6]{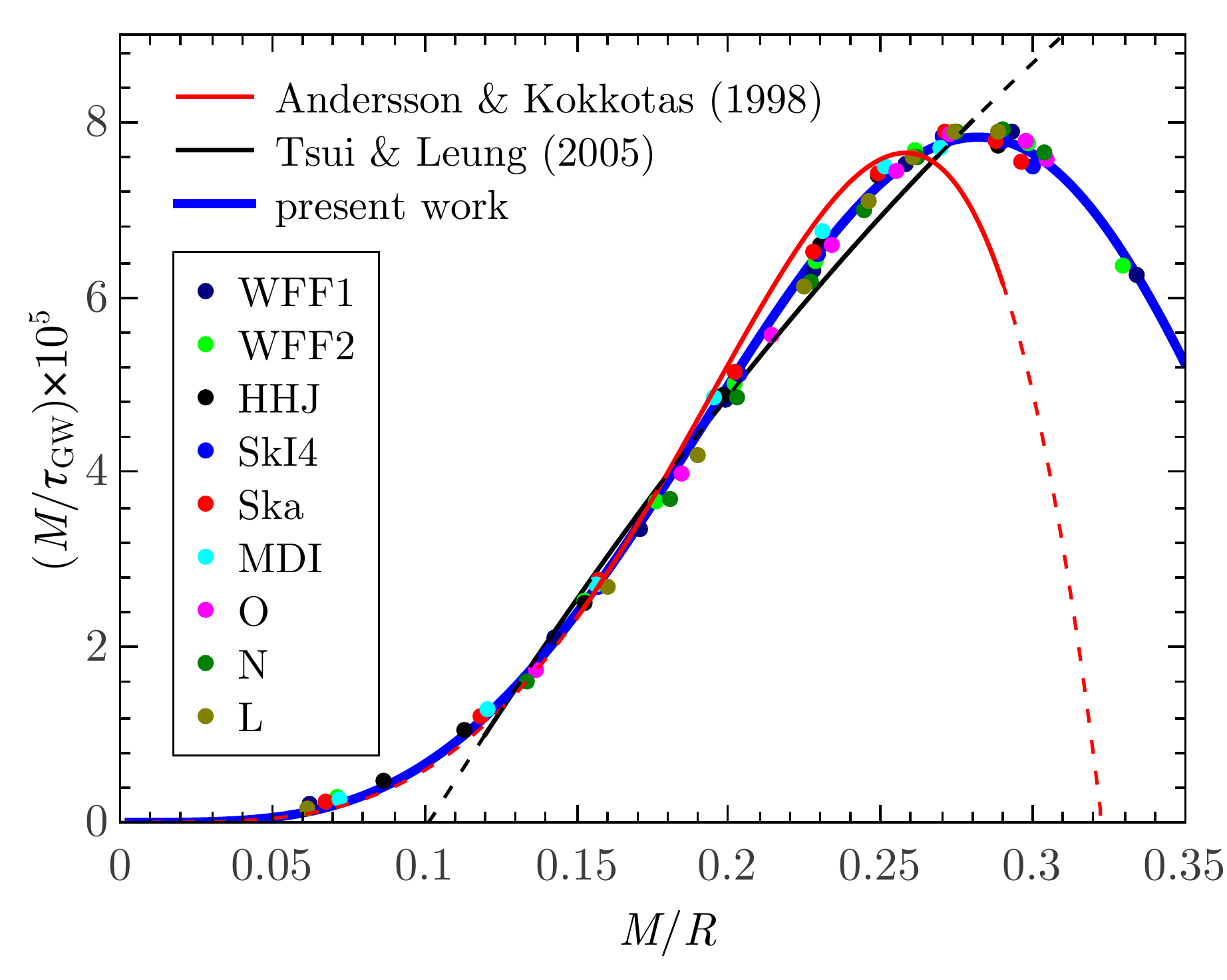}
\caption{Universal relation, Eq. (\ref{uniMnew}),  for the gravitational-wave damping timescale
 (shown here as $M/\tau_{\rm GW}$ vs. compactness, blue line). The previous relations by Andersson \& Kokkotas (1998), Eq.  (\ref{uniAKM}), and by Tsui \& Leung (2005), Eq. (\ref{uniTL}), are also shown with solid lines in their  respective range of validity and with dashed lines outside of that range.}
\label{fig:uniMoverTau}       
\end{figure}

  Andersson and Kokkotas  \cite{AnKok1998} showed that the damping timescale $\tau_{\rm GW}$ for $f-$modes, when scaled
by $M^3/R^4$, satisfies the following empirical relation, as a function of the compactness $M/R$ 
\begin{equation}
\tau_{\rm GW}\times M^3/R^4 = \frac{1}{0.0862-0.2672(M/R)},
\label{uniAK}
\end{equation}
(written here in our chosen units).
The expected   $M^3/R^4$ scaling follows from the lowest-order dependence of the ratio ${E^0_{\rm mode}}/ {<dE/dt>_{\rm GW}}$ on $M$ and $R$. It is obvious that Eq. (\ref{uniAK}) can also be written as 
\begin{equation}
M/\tau_{\rm GW} = 0.0862 (M/R)^4-0.2672(M/R)^5.
\label{uniAKM}
\end{equation}

A very different empirical relation for $\tau_{\rm GW}$ was proposed in  \cite{TsLe2005}, in order for both the real and imaginary parts of the eigenfrequency to satisfy the same complex relation. The empirical relation in  \cite{TsLe2005} was 
\begin{equation}
M/\tau_{\rm GW} =-6.2\times 10^{-5}+ 6.7\times10^{-4} (M/R)-5.8\times 10^{-4}(M/R)^2.
\label{uniTL}
\end{equation}In addition to the above relations, which depend only on $M$ and $R$, a highly accurate empirical relation was shown to exist in \cite{LaLeLi2010}, which, however, is a function of
an effective compatness, $\eta$, that involves the moment of inertia $I,$
see also \cite{DoKo2015,ChSoKa2015}.

Here, we extend the relation found by Andersson \& Kokkotas \cite{AnKok1998} to higher
order. We find that the numerical data for $\tau_{\rm GW}$\ obtained by solving directly for the  complex eigenvalues of the perturbation equations presented  in Section \ref{sec:lin}. are described very accurately by the following universal relation

\begin{equation}
M/\tau_{\rm GW} = 0.112(M/R)^4-0.53(M/R)^5+0.628(M/R)^6.
\label{uniMnew}
\end{equation}

This relation was constructed by taking into account only models with $M/R>0.1$. The reason behind this choice is that these models are physically interesting and of practical use. Yet, one should note that even in the 
region $M/R<0.1$ the agreement with the numerical data remains excellent.

Fig. \ref{fig:uniMoverTau} presents our numerical data. As $M/R$ increases, the configuration becomes more relativistic, so gravitational radiation
becomes a more effective damping mechanism, thus the damping time $\tau_{\rm GW}$ decreases. However, when $M/R$ becomes large enough, the peak of the exterior
potential emerges, making the situation analogous to barrier penetration in wave mechanics \cite{Det1975,AndKojKok1996}. Gravitational waves find it harder to escape to infinity
and thus $\tau_{\rm GW}$ becomes larger. According to our results, this occurs at roughly $M/R\approx0.28$. Combined with how $M$ behaves as $M/R$ increases, 
apparent in Fig. \ref{fig:1}, the shape of $M/\tau_{\rm GW}$ in Fig. \ref{fig:uniMoverTau} is explained.

Furthermore, Fig. \ref{fig:uniMoverTau} compares 
our new relation (\ref{uniMnew}) to the numerical data, as well as to the previously suggested relations (\ref{uniAKM}) and (\ref{uniTL}). It is evident that (\ref{uniAKM}) and (\ref{uniTL}) have comparable absolute errors in the limited range of $0.13<M/R<0.27$. Eq. (\ref{uniAKM}) describes fairly well the Newtonian limit, where $1/\tau_{GW}\rightarrow 0$ as $M/R\rightarrow 0$, but loses accuracy for very compact stars of $M/R>0.27$.  In contrast, relation (\ref{uniTL}) does not display the correct Newtonian limit and also loses accuracy for very compact stars. Our new relation (\ref{uniMnew}) describes the data with high accuracy for any value of the compactness, up to the most compact stars with $M/R=0.33$ that we consider here.

The fact that our new universal relation (\ref{uniMnew}) is highly accurate in the whole range of $0<M/R<0.33$ (the standard
statistical correlation coefficient is $0.9997$) eliminates (at least for neutron stars) the need for using
the moment of inertia, as was done in  \cite{LaLeLi2010,DoKo2015,ChSoKa2015}. The gravitational-wave damping timescale $\tau_{\rm
GW}$ can be thus described by a highly accurate universal relation that only
involves $M$ and $R$, for which the prospects of determination in gravitational-wave sources are realistic, whereas the simultaneous determination of the moment of inertia appears far less likely. For practical purposes, (\ref{uniMnew})
is thus a better choice over previously suggested relations. 
 
 \section{Accuracy of different approximations for the damping timescale}
 
\begin{figure}[]
\begin{subfigure}{\textwidth}
  \centering
  \includegraphics[width=0.80\linewidth]{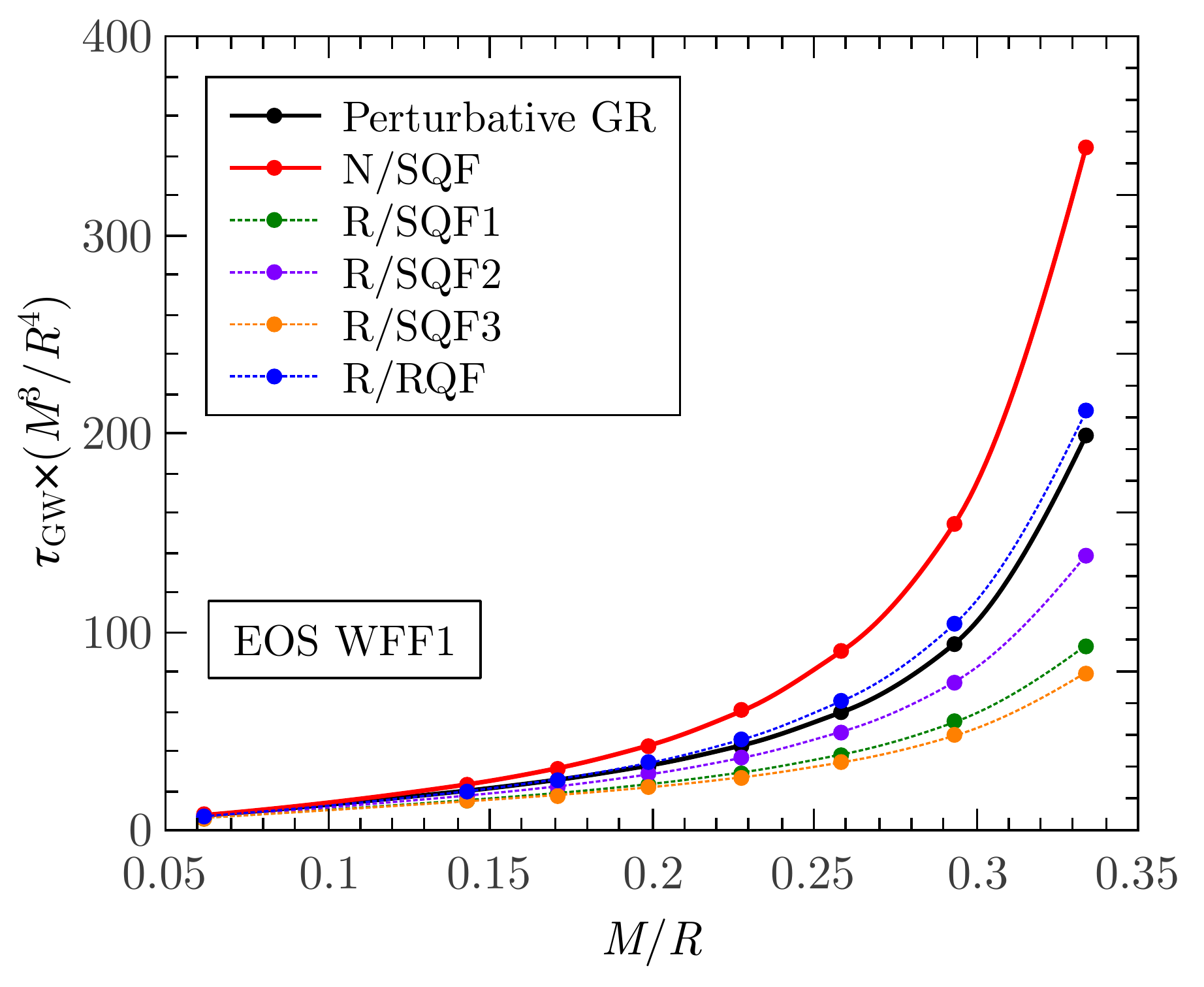}
  \caption{(\ref{sfig1})}
  \label{sfig1}
\end{subfigure}\\
\begin{subfigure}{\textwidth}
  \centering
  \includegraphics[width=0.80\linewidth]{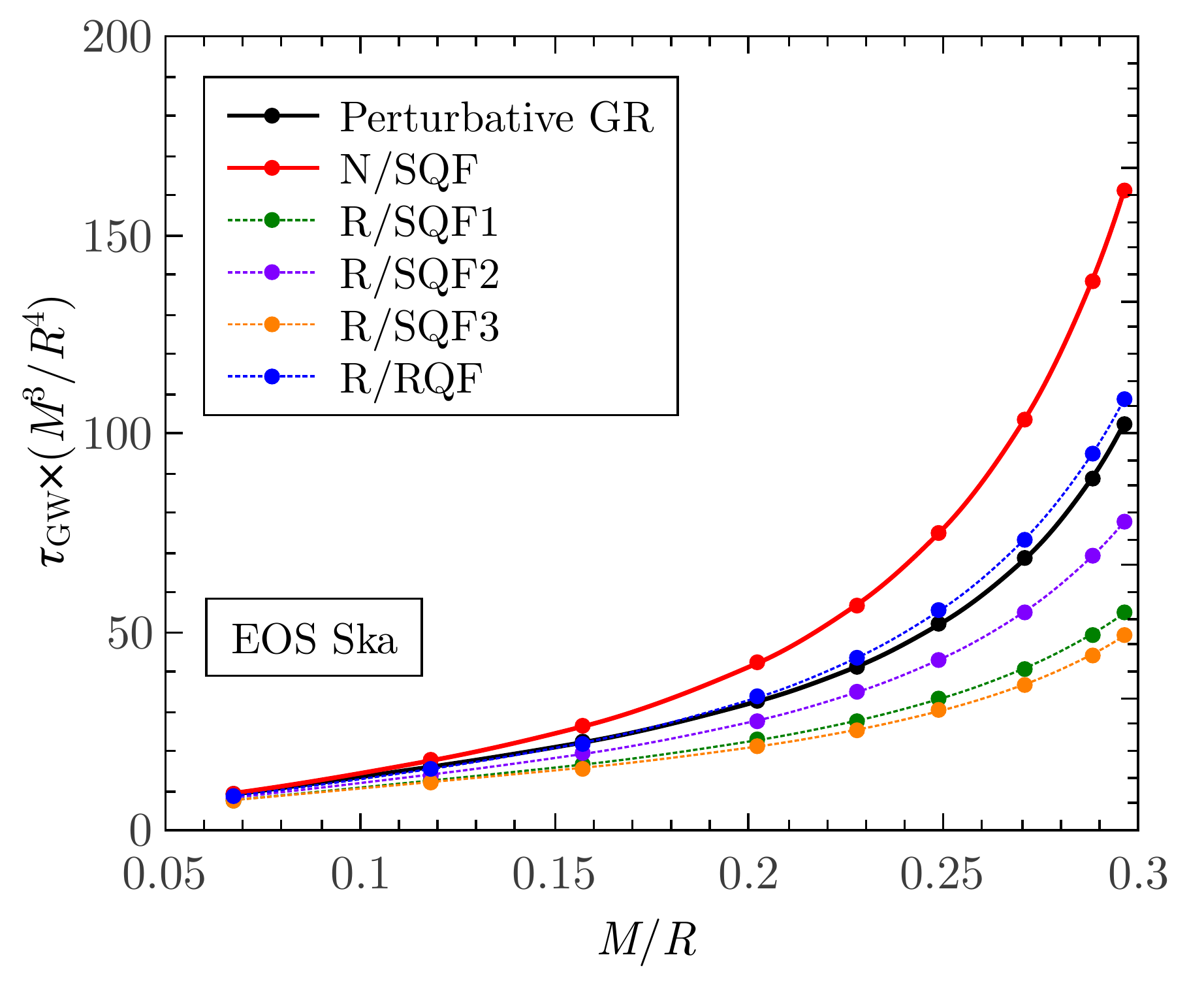}
  \caption{(\ref{sfig2})}
  \label{sfig2}
\end{subfigure}
\caption{Damping timescales $\tau_{\rm GW}$ computed through perturbation theory in full GR as well as through several different approximation methods (see text for definitions). Top panel (\ref{sfig1}): EOS WFF1, bottom panel (\ref{sfig2}): EOS Ska.}
\label{sfigg1}
\end{figure}

\begin{figure}[]
\begin{subfigure}{\textwidth}
  \centering
  \includegraphics[width=.80\linewidth]{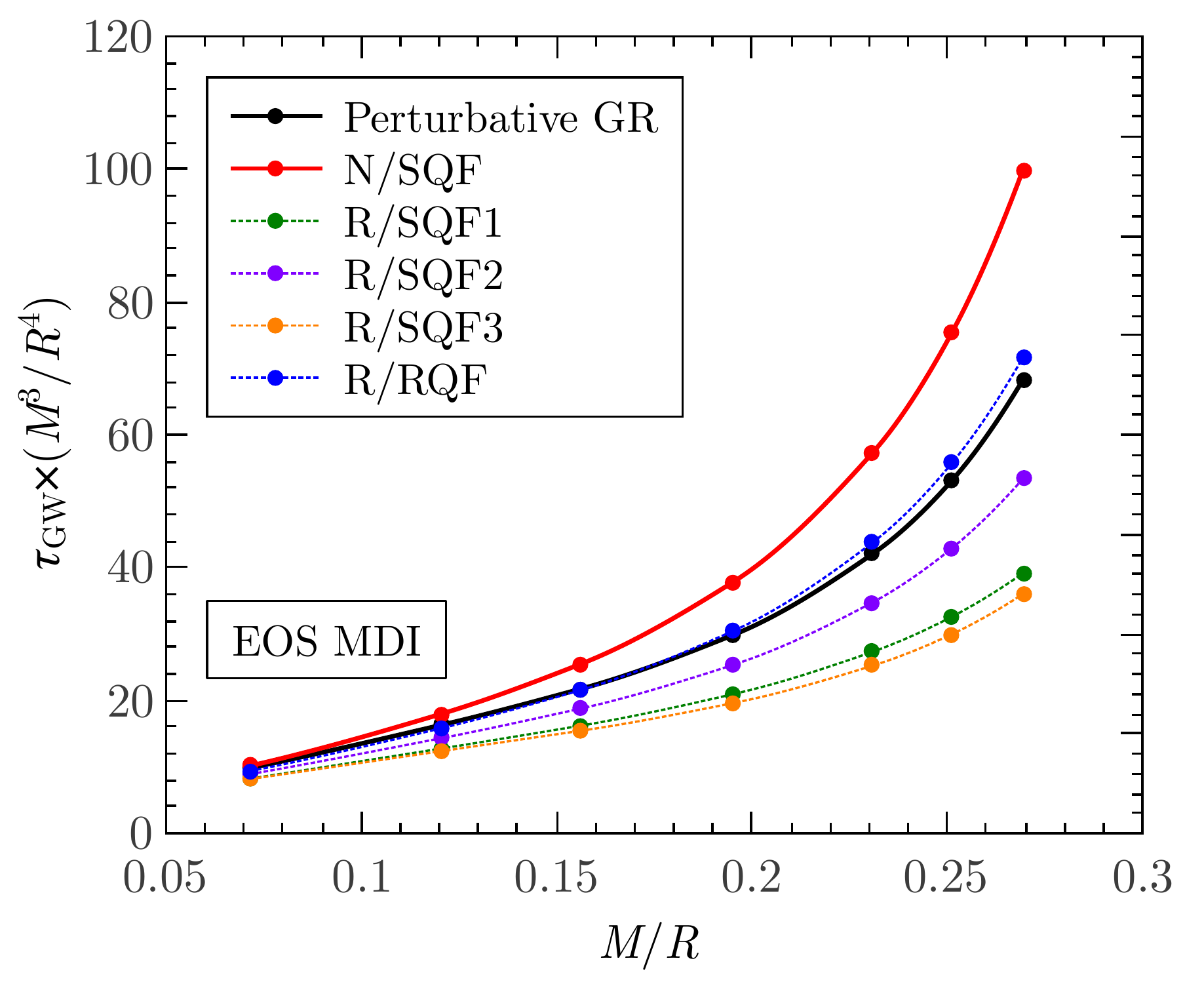}
  \caption{(\ref{sfig3})}
  \label{sfig3}
\end{subfigure} \\
\begin{subfigure}{\textwidth}
  \centering
  \includegraphics[width=.80\linewidth]{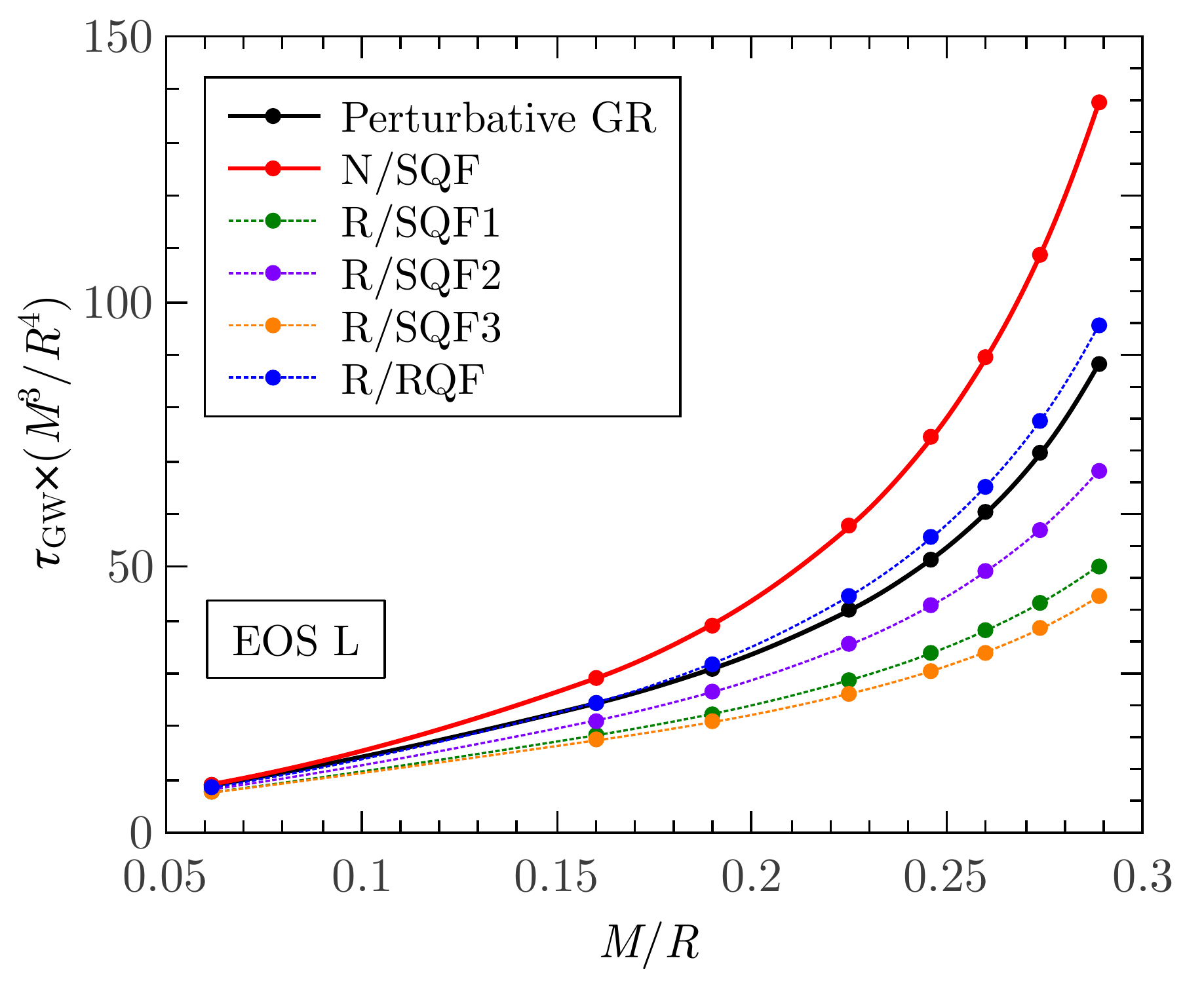}
  \caption{(\ref{sfig4})}
  \label{sfig4}
\end{subfigure}
\caption{Same as Fig. \ref{sfigg1}, but for EOS MDI (\ref{sfig3}) and L  (\ref{sfig4}).}
\label{sfigg2}
\end{figure}

 Approximate gravitational-wave damping timescales $\tau_{\rm GW}$ for the $l=2,$ $m=0$ $f-$modes, for all nine EOS and for different compactness $M/R$ (up to the maximum compactness reached by each EOS) were computed using the five different methods listed in Table \ref{tabb2}.
The different panels in Figs.  \ref{sfigg1} and  \ref{sfigg2}
show $\tau_{\rm GW}$, as obtained by the different approximation methods, in comparison to $\tau_{\rm GW}$ obtained by solving for the complex
eigenvalues of the system of perturbation equations described in Section
\ref{sec:lin} (only four representative EOS are shown, with results for the other five EOS being similar).
To our knowledge, 
this is the first such direct comparison of approximations for $\tau_{\rm GW}$ to the actual perturbative results, except for the standard quadrupole formula (N/SQF), which was compared to the actual perturbative result for polytropic models in \cite{BalDetLinSch1985}.

  \begin{figure}[h!]
  \centering
  \includegraphics[width=.80\linewidth]{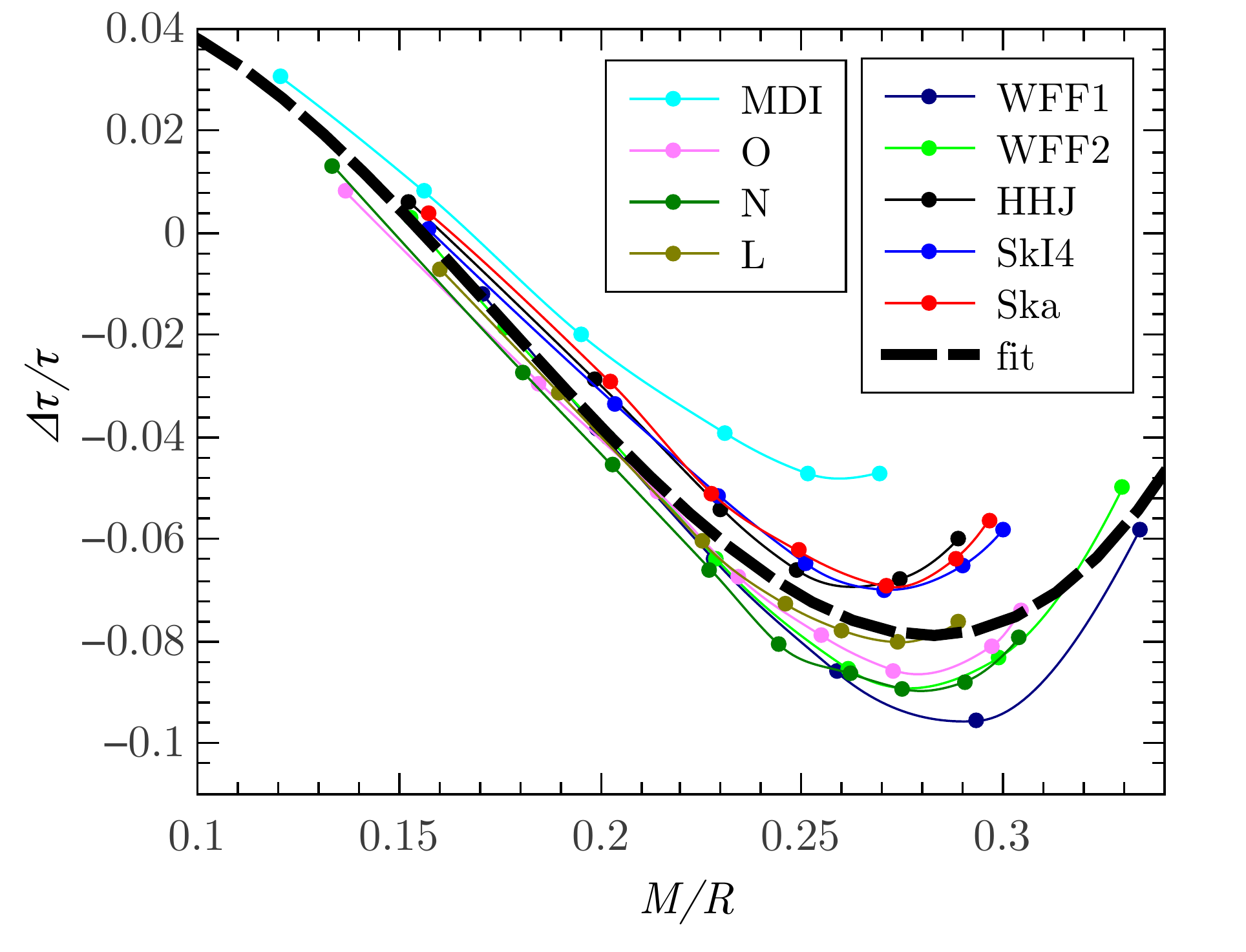}
\caption{The relative error $\Delta\tau/\tau$ vs. $M/R$ between the R/RQF approximation and perturbative GR for all
EOS under consideration. The universal 3rd-order fit is used for constructing the empirically-corrected relation (\ref{taucorrected}).}
\label{fig:3}      
\end{figure}

The detailed comparisons displayed in Figs.  \ref{sfigg1} and  \ref{sfigg2}
reveal that the R/RQF approximation (that is, combining the relativistic pulsation energy (\ref{enerns}) with a quadrupole formula for the luminosity, in which the density is replaced by the energy density) is by far the best of all five approximation methods considered here. For all EOS, the damping timescale  $\tau_{\rm
GW}$ computed with R/RQF is close to the relativistic perturbative result, with the maximum relative error being in the range $5\% - 10\%$, depending on the EOS. This is significantly more accurate than either the standard quadrupole formula N/SQF or the other approximations considered here (see below). R/RQF overestimates (underestimates) the perturbative results for compactness $\gtrsim 0.15$ ($\lesssim0.15$). The error does not depend monotonically on the stiffness of the EOS. Notice that for a given EOS the maximum error is attained for compactness $\sim 0.27-0.29$, while the models with maximum compactness have somewhat smaller error (see Fig. \ref{fig:3} below). Especially for mildly relativistic stars with compactness  $M/R \lesssim\ 0.2$, the R/RQF approximation has a relative error smaller than $6\%$ for all EOS in our sample. 

The other approximations considered here have significantly larger relative errors, that reach up to 50\%, 30\% and 63\% for  R/SQF1, R/SQF2 and R/SQF3, respectively (in all cases these approximations
underestimate the perturbative result). \ These largest errors are attained for the softest EOS in our sample (WFF1) and for models with the highest compactness of $\sim 0.34$. Even for mildly relativistic stars, the relative error when using the approximations R/SQF1, R/SQF1 and R/SQF3 is still significant. The standard quadrupole formula N/SQF overestimates the perturbative results, with the error increasing as the compactness increases (in the Newtonian limit we correctly recover the perturbative results). The maximum error with N/SQF is 73\% for the most compact model of EOS WFF1.

One should keep in mind that \eqref{enerns} is exact for spherical stars. As such, it is bound to be a better choice than its Newtonian version 
\eqref{pulsnewt}. In the meantime, we have shown numerically that R/RQF is a better approximation than any of the R/SQF1,R/SQF2 and R/SQF3. This 
is a result that could be further studied through a post-Newtonian analysis. In addition, in the case of rotating stars, \eqref{enerns1} and 
\eqref{quadformula} could be employed in order to compute $\tau_{\rm GW}$ as long as the real part of the eigenfunctions is known (see, e.g. \cite{SterFried1998,MorSterBla1999,Yosh2012} for computations of the real part of the eigenfunction of perturbations in rapidly rotating neutron star models).

We define the relative difference between the actual perturbative result $\tau_{\rm GW}^{\rm pert}$ and an approximate result $\tau_{\rm GW}^{\rm approx}$ (obtained with one of the approximate methods listed in Table \ref{tabb2}) as
\begin{equation}
\frac{\Delta \tau}{\tau}  = \left (\frac{\tau_{\rm GW}^{\rm pert}-\tau_{\rm GW}^{\rm approx}}{\tau_{\rm
GW}^{\rm approx}} \right).
\label{Deltatau}
\end{equation}
Fig. \ref{fig:3} displays the relative difference $\Delta \tau /\tau$ as a function of compactness $M/R$ for the most accurate approximation R/RQF, shown for all nine EOS.
The relative difference vanishes for compactness $M/R\sim 0.15$ and increases for smaller or larger compactness, reaching a maximum magnitude close to the maximum compactness allowed by each EOS (for the most compact stars, the approximate relation becomes more accurate, again). The maximum relative difference, as defined by (\ref{Deltatau}), is 4.5\% for the MDI EOS (which has the smallest maximum mass), 7\% for the EOS HHJ, SkI4 and Ska, which are of intermediate stiffness and 8\%-9.5\% for the very soft EOS WFF1, WFF2 as well as for the very stiff EOS N and L. 

We thus find that t\textit{he approximate formula R/RQF yields gravitational-wave damping times that are within 10\% of their exact value}, which is\ significantly more accurate than either the standard quadrupole formula N/SQF or the other three modifications R/SQF1, R/SQF2 and R/SQF3.

 \section{Empirically corrected approximate relation}
 \label{sec:emp}
 
  \begin{figure}[h!]
  \centering
  \includegraphics[width=0.80\linewidth]{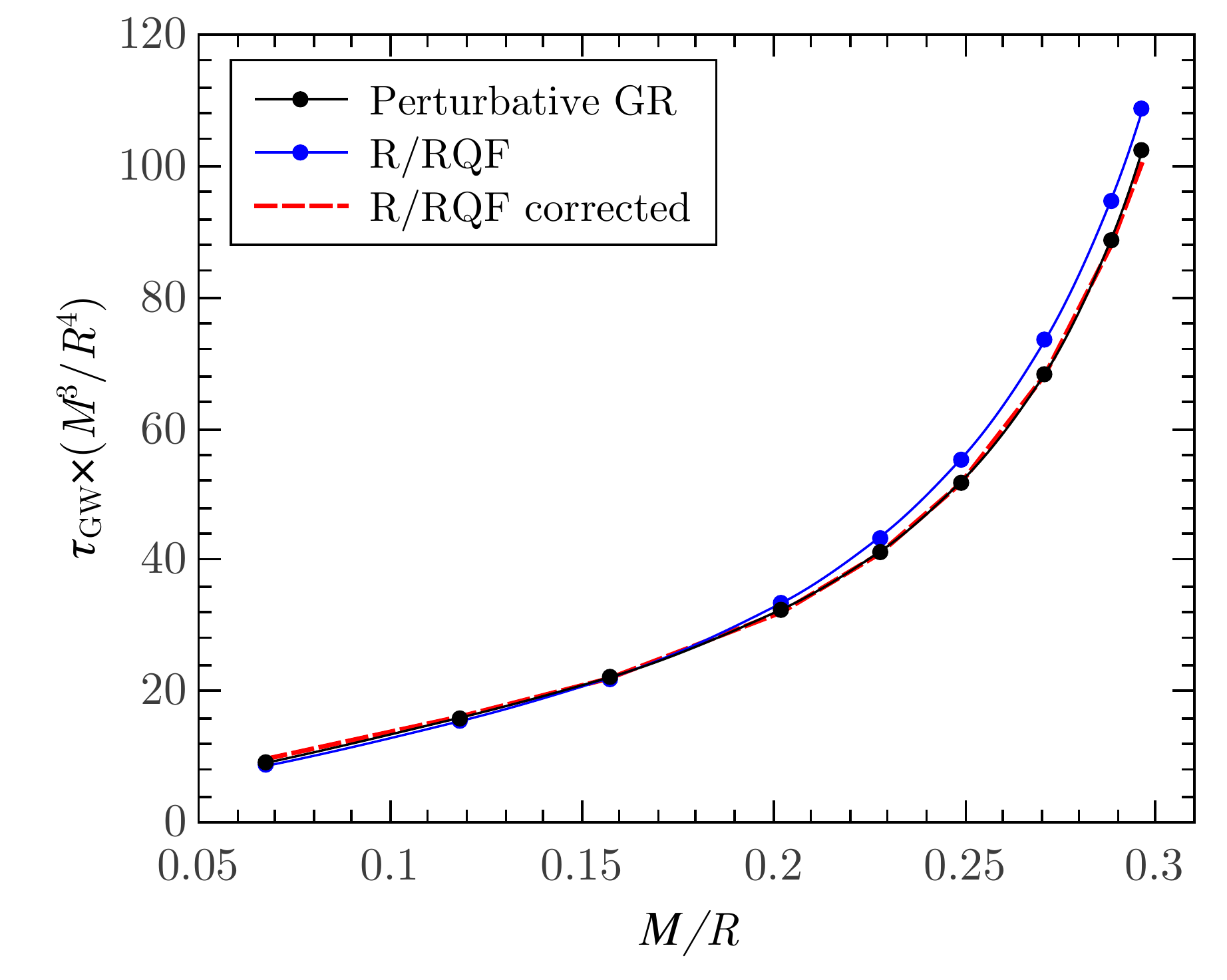}
\caption{Gravitational-wave damping timescale $\tau_{\rm GW}$ for quadrupole oscillations, computed with perturbation
theory in full GR, the R/RQF approximation  and the empirically corrected relation (\ref{taucorrected}). The latter reproduces the perturbative results with high accuracy.}
\label{fig:5}       
\end{figure}

 We find that the relative difference between the R/RQF approximation and the perturbative results, as defined in (\ref{Deltatau}) is described well by the following universal, 3rd-order empirical relation
 \begin{equation}
\Delta\tau/\tau \simeq 1.47 \left(\frac{M}{R}\right) - 
13.3 \left(\frac{M}{R}\right)^2 + 25.3 \left(\frac{M}{R}\right)^3,
\label{Deltataufit} 
\end{equation}
 (see Fig. \ref{fig:3}). In constructing (\ref{Deltataufit}), we only used models with $M/R>0.1$ and restricted the fit to satisfy the constraint $\Delta \tau/\tau \rightarrow 0$ as $M/R\rightarrow 0$. 

Solving (\ref{Deltatau}) for $\tau_{\rm GW}^{\rm pert} =\tau_{\rm
GW}^{\rm approx} \left( 1 + \Delta\tau \right/\tau)$ we thus obtain an \textit{empirically corrected }damping timescale \begin{equation}
   \tau_{\rm GW} \simeq \tau_{\rm
GW}^{\rm (R/RQF)}\left[1+1.47 \left(\frac{M}{R}\right) - 
13.3 \left(\frac{M}{R}\right)^2 + 25.3 \left(\frac{M}{R}\right)^3\right],
\label{taucorrected}
 \end{equation}
 where the correction factor in brackets is EOS-independent.
The maximum relative error of this relation is only 3\%.
An example of the empirically corrected damping timescale, as a function of compactness, is shown in Fig. \ref{fig:5}, in the case of EOS Ska.
  
 Instead of using  the empirical relation (\ref{Deltataufit}) in (\ref{taucorrected}), valid for all EOS, one may wish to  construct individual corrected relations using
a 3-order fit for $\Delta \tau/\tau$   for each EOS separately. Table \ref{tab2}, lists the corresponding fits for each EOS and the corresponding maximum relative error in  $\tau_{\rm GW}$ in each case.  
 \begin{table}[h!]
\begin{center}
 {
\renewcommand{\arraystretch}{1.5}
\caption{Empirical fits to the relative error $\Delta\tau/\tau$ for each individual EOS. The last column shows the maximum relative error of (\ref{taucorrected}) when these fits are used in place of the universal relation (\ref{Deltataufit}).}
\label{tab2}      
\begin{tabular}{clc}
\hline\noalign{\smallskip}
EOS & \multicolumn{1}{c}{Cubic Fit} & Maximum relative error in $\tau_{\textrm GW}$  \\
\noalign{\smallskip}\hline\noalign{\smallskip}
WFF1 & $\left.\frac{\Delta\tau}{\tau}\right|_{WFF1}= 1.79 \left(\frac{M}{R}\right) -  16.1 \left(\frac{M}{R}\right)^2 + 30.5 \left(\frac{M}{R}\right)^3$ & $0.66\%$\\ 
WFF2 & $\left.\frac{\Delta\tau}{\tau}\right|_{WFF2}= 1.81 \left(\frac{M}{R}\right) -  16.4 \left(\frac{M}{R}\right)^2 + 31.8 \left(\frac{M}{R}\right)^3$ & $0.39\%$\\
HHJ & $\left.\frac{\Delta\tau}{\tau}\right|_{HHJ}= 1.6 \left(\frac{M}{R}\right) -  14.6 \left(\frac{M}{R}\right)^2 + 28.7 \left(\frac{M}{R}\right)^3$ & $0.41\%$\\
SkI4 & $\left.\frac{\Delta\tau}{\tau}\right|_{SkI4}= 1.5 \left(\frac{M}{R}\right) -  13.7 \left(\frac{M}{R}\right)^2 + 26.6 \left(\frac{M}{R}\right)^3$ & $0.34\%$\\
Ska & $\left.\frac{\Delta\tau}{\tau}\right|_{Ska}= 1.62 \left(\frac{M}{R}\right) -  14.6 \left(\frac{M}{R}\right)^2 + 28.6 \left(\frac{M}{R}\right)^3$ & $0.44\%$\\
MDI & $\left.\frac{\Delta\tau}{\tau}\right|_{MDI}= 1.43 \left(\frac{M}{R}\right) -  12.8 \left(\frac{M}{R}\right)^2 + 25.4 \left(\frac{M}{R}\right)^3$ & $0.11\%$\\
O & $\left.\frac{\Delta\tau}{\tau}\right|_{O}= 1.39 \left(\frac{M}{R}\right) -  13.1 \left(\frac{M}{R}\right)^2 + 25.2 \left(\frac{M}{R}\right)^3$ & $0.43\%$\\
N & $\left.\frac{\Delta\tau}{\tau}\right|_{N}= 1.55 \left(\frac{M}{R}\right) -  14.5 \left(\frac{M}{R}\right)^2 + 28 \left(\frac{M}{R}\right)^3$ & $0.30\%$\\
L & $\left.\frac{\Delta\tau}{\tau}\right|_{L}= 1.55 \left(\frac{M}{R}\right) - 14.5 \left(\frac{M}{R}\right)^2 + 28.2 \left(\frac{M}{R}\right)^3$ & $0.15\%$\\
\noalign{\smallskip}\hline
\end{tabular}
}
\end{center}
\end{table}

  \section{Discussion and outlook}
  
Existing estimates of the gravitational-wave damping timescale  $\tau_{\rm
GW}$ of the dominant
quadrupole oscillation mode in the case of rapidly rotating stars are based on using
a Newtonian estimate for the energy of the mode, in combination with
the lowest-order post-Newtonian quadrupole formula for estimating the gravitational-wave
luminosity (e.g. \cite{DoGaKoKr2013,DoKo2015}).
We investigated a number of other choices for estimating $\tau_{\rm
GW}$ (in the nonrotating limit) and compared their relative accuracy
with respect to the perturbative result in full general relativity. We found that a
specific choice, which we call R/RQF and which was used in \cite{Burg2011}, stands
out for its higher accuracy. 
Furthermore, we found an EOS-independent empirical relation that corrects for the difference
between the approximate and exact results, leading to a empirically corrected formula for $\tau_{\rm
GW}$ that has a maximum relative error of only 3\%. The expressions involved in the R/RQF are sufficiently general
to be  applied also for rapidly
rotating stars. 

Investigating the EOS-dependence of damping timescales of quadrupole oscillations for nonrotating stars, we were also able to extend the empirical relation first found in  \cite{AnKok1998} to higher order. Our new expression is more accurate for models of high compactness and is superior to a different relation proposed in \cite{TsLe2005}.  In addition, it eliminates (at least for neutron stars)\ the need for using a rescaled, effective compatness, that involves, apart from mass and radius, also the moment of inertia $I$, as was proposed and used in \cite{LaLeLi2010,DoKo2015,ChSoKa2015}. Our new high-order empirical relation could also be extended to rotating stars in a way that only
 $M$ and $R$ and a third parameter describing the amount of rotation is involved.

We plan to revisit the generalizations to rotating stars in future work.

\begin{acknowledgements}
We would like to thank Ch. Moustakidis for providing us data for the EOS MDI, SkI4, Ska and HHJ. We are grateful to Andreas Bauswein, Daniela Doneva, John Friedman and Kostas Kokkotas for useful discussions and for comments on the manuscript. Partial support comes from the COST action PHAROS\  (CA16214) and the DAAD Germany-Greece grant ID 57340132.
\end{acknowledgements}


\bibliographystyle{spmpsci}  
\bibliography{references}

\end{document}